\documentclass[12pt,a4paper]{article}

\usepackage{ifthen} 
\newboolean{articletitles}
\setboolean{articletitles}{true} 
\newboolean{uprightparticles}
\setboolean{uprightparticles}{false} 
\newboolean{inbibliography}
\setboolean{inbibliography}{false} 

\usepackage{microtype}
\usepackage{lineno}  
\usepackage{xspace} 

\usepackage{graphicx}  
\usepackage{color}
\usepackage{colortbl}
\graphicspath{{./figs/}} 

\usepackage{amsmath} 
\usepackage{amssymb}
\usepackage{amsfonts}
\usepackage{mathrsfs,bm,url,color}
\usepackage{upgreek} 
\usepackage{blindtext} 

\newcommand*\patchAmsMathEnvironmentForLineno[1]{%
\expandafter\let\csname old#1\expandafter\endcsname\csname #1\endcsname
\expandafter\let\csname oldend#1\expandafter\endcsname\csname
end#1\endcsname
 \renewenvironment{#1}%
   {\linenomath\csname old#1\endcsname}%
   {\csname oldend#1\endcsname\endlinenomath}%
}
\newcommand*\patchBothAmsMathEnvironmentsForLineno[1]{%
  \patchAmsMathEnvironmentForLineno{#1}%
  \patchAmsMathEnvironmentForLineno{#1*}%
}
\AtBeginDocument{%
\patchBothAmsMathEnvironmentsForLineno{equation}%
\patchBothAmsMathEnvironmentsForLineno{align}%
\patchBothAmsMathEnvironmentsForLineno{flalign}%
\patchBothAmsMathEnvironmentsForLineno{alignat}%
\patchBothAmsMathEnvironmentsForLineno{gather}%
\patchBothAmsMathEnvironmentsForLineno{multline}%
}

\usepackage{hyperref}    
\usepackage[all]{hypcap} 

\def\ux85 {\mbox{UX85}\xspace}

\ifthenelse{\boolean{uprightparticles}}%
{

 \def\PDelta      {\ensuremath{\Delta}\xspace}                 
 \def\PXi      {\ensuremath{\Xi}\xspace}                 
 \def\PLambda      {\ensuremath{\Lambda}\xspace}                 
 \def\PSigma      {\ensuremath{\Sigma}\xspace}                 
 \def\POmega      {\ensuremath{\Omega}\xspace}                 
 \def\PUpsilon      {\ensuremath{\Upsilon}\xspace}                 
 
 \def\PB      {\ensuremath{\mathrm{B}}\xspace}                 
                  
 \def\PD      {\ensuremath{\mathrm{D}}\xspace}

 \def\PK      {\ensuremath{\mathrm{K}}\xspace}

 \def\Pb      {\ensuremath{\mathrm{b}}\xspace}                 
 \def\Pc      {\ensuremath{\mathrm{c}}\xspace}

 \def\Pi      {\ensuremath{\mathrm{i}}\xspace}

}
{

 \mathchardef\PDelta="7101
 \mathchardef\PXi="7104
 \mathchardef\PLambda="7103
 \mathchardef\PSigma="7106
 \mathchardef\POmega="710A
 \mathchardef\PUpsilon="7107
                  
 \def\PB      {\ensuremath{B}\xspace}                 
                  
 \def\PD      {\ensuremath{D}\xspace}

 \def\PK      {\ensuremath{K}\xspace}

 \def\Pb      {\ensuremath{b}\xspace}                 
 \def\Pc      {\ensuremath{c}\xspace}

 \def\Pi      {\ensuremath{i}\xspace}

}

\def\cquark    {\ensuremath{\Pc}\xspace}

\def\bquark    {\ensuremath{\Pb}\xspace}

\def\kaon  {\ensuremath{\PK}\xspace}
  \def\Kbar  {\kern 0.2em\overline{\kern -0.2em \PK}{}\xspace}

\def\Kz    {\ensuremath{\kaon^0}\xspace}
\def\Kzb   {\ensuremath{\Kbar^0}\xspace}
\def\KzKzb {\ensuremath{\Kz \kern -0.16em \Kzb}\xspace}
\def\Kp    {\ensuremath{\kaon^+}\xspace}
\def\Km    {\ensuremath{\kaon^-}\xspace}

\def\KpKm  {\ensuremath{\Kp \kern -0.16em \Km}\xspace}

\def\Dbar    {\kern 0.2em\overline{\kern -0.2em \PD}{}\xspace}
\def\D       {\ensuremath{\PD}\xspace}

\def\Dz      {\ensuremath{\D^0}\xspace}
\def\Dzb     {\ensuremath{\Dbar^0}\xspace}
\def\DzDzb   {\ensuremath{\Dz {\kern -0.16em \Dzb}}\xspace}
\def\Dp      {\ensuremath{\D^+}\xspace}
\def\Dm      {\ensuremath{\D^-}\xspace}

\def\DpDm    {\ensuremath{\Dp {\kern -0.16em \Dm}}\xspace}

\def\Bbar    {\ensuremath{\kern 0.18em\overline{\kern -0.18em \PB}{}}\xspace}

  \def\Y#1S{\ensuremath{\PUpsilon{(#1S)}}\xspace}

\def\Lbar {\ensuremath{\kern 0.1em\overline{\kern -0.1em\PLambda}}\xspace}

\def\to                 {\ensuremath{\rightarrow}\xspace}

\def\CP                {\ensuremath{C\!P}\xspace}

\def\AT#1     {\ensuremath{A_{\mathrm{T}}^{#1}}\xspace}           

\def\C#1      {\ensuremath{\mathcal{C}_{#1}}\xspace}                       
\def\Cp#1     {\ensuremath{\mathcal{C}_{#1}^{'}}\xspace}                    
\def\Ceff#1   {\ensuremath{\mathcal{C}_{#1}^{\mathrm{(eff)}}}\xspace}        
\def\Cpeff#1  {\ensuremath{\mathcal{C}_{#1}^{'\mathrm{(eff)}}}\xspace}       
\def\Ope#1    {\ensuremath{\mathcal{O}_{#1}}\xspace}                       
\def\Opep#1   {\ensuremath{\mathcal{O}_{#1}^{'}}\xspace}                    

\newcommand{\tev}{\ensuremath{\mathrm{\,Te\kern -0.1em V}}\xspace}
\newcommand{\gev}{\ensuremath{\mathrm{\,Ge\kern -0.1em V}}\xspace}
\newcommand{\mev}{\ensuremath{\mathrm{\,Me\kern -0.1em V}}\xspace}
\newcommand{\kev}{\ensuremath{\mathrm{\,ke\kern -0.1em V}}\xspace}
\newcommand{\ev}{\ensuremath{\mathrm{\,e\kern -0.1em V}}\xspace}
\newcommand{\gevc}{\ensuremath{{\mathrm{\,Ge\kern -0.1em V\!/}c}}\xspace}
\newcommand{\mevc}{\ensuremath{{\mathrm{\,Me\kern -0.1em V\!/}c}}\xspace}
\newcommand{\gevcc}{\ensuremath{{\mathrm{\,Ge\kern -0.1em V\!/}c^2}}\xspace}
\newcommand{\gevgevcccc}{\ensuremath{{\mathrm{\,Ge\kern -0.1em V^2\!/}c^4}}\xspace}
\newcommand{\mevcc}{\ensuremath{{\mathrm{\,Me\kern -0.1em V\!/}c^2}}\xspace}

\def\mub{\ensuremath{\rm \,\upmu b}\xspace}

\def\gsim{{~\raise.15em\hbox{$>$}\kern-.85em
          \lower.35em\hbox{$\sim$}~}\xspace}
\def\lsim{{~\raise.15em\hbox{$<$}\kern-.85em
          \lower.35em\hbox{$\sim$}~}\xspace}

\def\tell1  {TELL1\xspace}
\def\ukl1   {UKL1\xspace}

\usepackage{cite} 
\usepackage{mciteplus}
\usepackage{overpic}
\usepackage{subfigure}

\usepackage[noabbrev,capitalise]{cleveref}

\usepackage{floatrow}

\newfloatcommand{capbtabbox}{table}[][\FBwidth]

\usepackage{caption}
\usepackage{longtable}
\usepackage{ifpdf}

\newcommand{\re}[2][()] {\ifthenelse{\equal{#1}{()}}{{\ensuremath{{\rm \, Re}}\left(#2\right)}}
                                                    {{\ensuremath{{\rm \, Re}}\left[#2\right]}}}
\newcommand{\im}[2][()] {\ifthenelse{\equal{#1}{()}}{{\ensuremath{{\rm \, Im}}\left(#2\right)}}
                                                    {{\ensuremath{{\rm \, Im}}\left[#2\right]}}}

\definecolor{orange}{rgb}{1,0.5,0}

\usepackage{booktabs}
\usepackage{rotating}
\usepackage{multirow}

\newcommand\snowmass{\begin{center}\rule[-0.2in]{\textwidth}{0.01in}\\\rule{\textwidth}{0.01in}\\
\vskip 0.1in Submitted to the Proceedings of the US Community Study\\ 
on the Future of Particle Physics (Snowmass 2021)\\ 
\rule{\textwidth}{0.01in}\\\rule[+0.2in]{\textwidth}{0.01in} \end{center}}

\begin{document}
\renewcommand{\thefootnote}{\fnsymbol{footnote}}
\setcounter{footnote}{1}
\begin{titlepage}

\snowmass
\vspace*{1.5cm}

{\bf\boldmath\huge
\begin{center}
\CP violation in \bquark and \cquark quark decays
\end{center}
}

\vspace*{0.5cm}

\begin{center}
Avital Dery$^{1}$,
Yuval Grossman$^{1}$,
Stefan Schacht$^{2}$,
Diego Tonelli$^{3}$
\bigskip\\
{\it\footnotesize
$^1$Department of Physics, LEPP, Cornell University, Ithaca, NY 14853, USA\\
$^2$Department of Physics and Astronomy, University of Manchester, Manchester M13 9PL, United Kingdom\\
$^3$INFN Sezione di Trieste, I-34127 Trieste, Italy
}
\end{center}

\vspace{\fill}

\begin{abstract}
Generically, non-Standard-Model particles contribute order-one \CP-violating phases to processes, as \CP is not a fundamental symmetry of nature. The exploration of \CP violation becomes therefore a sensitive search for non-Standard-Model physics. We briefly review the current status of \CP-violation studies in bottom- and charm-quark transitions, focusing on those quantities that are most sensitive to non-Standard-Model contributions, and discuss opportunities and challenges for the next decade and beyond.
\end{abstract}

\vspace{\fill}

\end{titlepage}

\pagestyle{empty}  


\newpage
\renewcommand{\thefootnote}{\arabic{footnote}}
\setcounter{footnote}{0}
\tableofcontents
\cleardoublepage
\pagestyle{plain} 
\setcounter{page}{1}
\pagenumbering{arabic}


\graphicspath{{figs/}}
\allowdisplaybreaks

\section{Introduction and overview}

Nonconservation of charge-parity symmetry (\CP violation) was unexpectedly discovered as a $10^{-3}$ effect in neutral kaon decays in 1964~\cite{Christenson:1964fg}. In 1973 an approach to accommodate economically \CP violation in the interactions of quarks was proposed~\cite{Kobayashi:1973fv}. However, since for many years the only experimental information on the phenomenon was limited to a single number restricted to the phenomenology of kaon mixing, various competing interpretations existed including those proposing new forces~\cite{Wolfenstein:1964ks}. Only in the 1990's observation of direct \CP violation in kaon decays indicated that the phenomenon was truly a property of the weak interactions~\cite{NA31:1988eyf, NA48:1999szy, KTeV:1999kad} and could therefore be studied in the dynamics of other hadrons. Indeed, since the 1980's theorists suggested that \CP-violating effects in bottom-meson transitions could have sizes much larger than in kaons~\cite{Bigi:1981qs, Carter:1980tk}. The propitious coincidences that (i) $B$ mesons undergo flavor oscillations; (ii) $B$ mesons have lifetimes sufficiently long to allow such oscillations be observed with then newly available silicon-based position-sensitive detectors; and (iii) $b$ quarks have a sizable coupling with $u$ quarks, thus allowing a truly $3\times3$ quark-family dynamics; made investigation of \CP violation in $B$ decays possible. The major ensuing breakthrough was the 2001 observation of large \CP violation in $B$ meson decays~\cite{BaBar:2001pki, Belle:2001zzw}, which conclusively established the Kobayashi-Maskawa (KM) phase~\cite{Kobayashi:1973fv} as the leading source of \CP violation in the Standard Model (SM). More recently \CP violation has been observed also in charm-meson decays~\cite{LHCb:2019hro} and evidence for \CP violation has been reported in $b$-baryon decays~\cite{LHCb:2016yco}.  Along the way, great progress has been achieved in determining the parameters of the Cabibbo-Kobayashi-Maskawa (CKM) matrix, which are now known with $10^{-2}$ accuracy. 

In recent years, emphasis of \CP violation studies has been shifting from confirming and establishing the KM picture, to using \CP violating observables as sensitive probes for a broad class of non-SM theories. Since \CP is not a fundamental symmetry of nature, non-SM particles are generically expected to contribute order-one \CP-violating phases to processes. Therefore, the exploration of \CP violation is naturally recast as a search for non-SM physics. For instance, recent semileptonic and rare $B$ decay data show various indications of lepton-flavor nonuniversality~\cite{Amhis:2019ckw}, suggesting deviations of the observations from the SM expectations.  If these ``anomalies" are hints of real effects, we could expect deviations from the SM also in $B$ decay \CP-violating asymmetries. Similar considerations~\cite{Gershon:2021pnc} apply to the anomalies found in $B$ decays governed by  $b\rightarrow c\bar{u} q$ transitions~\cite{Bordone:2020gao, Iguro:2020ndk, Fleischer:2021cwb}. Current experimental flavor physics enjoys the unprecedented opportunity of relying on two dedicated, state-of-the-art experiments, LHCb~\cite{LHCb:2008vvz,LHCb:2014set} and Belle~II~\cite{Belle-II:2010dht, Belle-II:2018jsg}, that operate simultaneously in complementary experimental environments that provide synergistic reach. The current and next generation of measurements aim at an order-of-magnitude improvement in sensitivity over the next decade. This, combined with equivalent advances in theory and lattice-QCD calculations, will probe SM extensions at energy scales much higher than those accessible in direct searches.

\subsection{Theory overview}

The theoretical treatment of \CP violation in charm and beauty decays has been covered in many excellent reviews, see \emph{e.g.},~Refs.~\cite{Nir:2020mgy, Nir:2020jtr, Silvestrini:2019sey, Zupan:2019uoi, Grossman:2017thq, Buras:2005xt}. It is not our aim here to repeat these overviews. Instead, in this introduction we want to touch upon recent important developments that are exciting right now in the context of the current anomalies and the foreseeable progress in the next few years.

For nonleptonic $B$ decays, the tensions of QCD factorization (QCDF) results \cite{Beneke:1999br,Beneke:2000ry, Beneke:2001ev} for $B\rightarrow DP$, $P=K,\pi$ with data make further systematic tests of factorization in non-leptonic decays an extremely interesting direction for future research.
Another interesting question, in the area of charmless $B$ decays, is the future of the so-called $K\pi$ puzzle. Commonly, this indicates early tensions of the data with an isospin-based sum-rule that relates several $B\rightarrow K\pi$ \CP asymmetries. The isospin sum rule can be formulated in such a way as to predict the least constrained involved \CP asymmetry, namely~\cite{LHCb:2020dpr}
\begin{align}
A_{CP}(B^0\rightarrow K^0\pi^0) &= -0.138 \pm 0.025 \qquad \text{(isospin sum rule prediction \cite{Gronau:2005kz,LHCb:2020dpr})\,,} \\
A_{CP}(B^0\rightarrow K^0\pi^0) &= \qquad 0.01 \pm 0.10  \qquad \text{(current exp. average~\cite{Amhis:2019ckw})\,.}
\end{align}
This corresponds at this time to a $1.4\sigma$ difference.  More precise measurements of the relevant \CP asymmetries are crucial to test the $K\pi$ isospin sum-rule further.
We  highlight the important progress on the extraction of $\gamma/\phi_3 \equiv \arg (-V_{ud}V^{*}_{ub}/V_{cd}V^{*}_{cb})$, especially in the latest analyses where charm-physics inputs and $\gamma$ are extracted within a unified framework~\cite{LHCb:2021dcr}. From the theory perspective, the special characteristic about $\gamma$ is that hadronic effects can be brought almost entirely under control, such that the ultimate precision for $\gamma$ extractions is quoted as $\lesssim \mathcal{O}(10^{-7})$~\cite{Brod:2013sga}, making $\gamma$ an exquisitely precise and reliable reference for SM \CP violation.\\

Important opportunities reside in charm dynamics. Charm \CP violation is a unique gate for probing the flavor structure of the up sector, and gives important constraints on several non-SM models, like 
Z' models \cite{Chala:2019fdb,Bause:2020obd}, 
Two-Higgs-Doublet models, and 
supersymmetric models \cite{Dery:2019ysp, Altmannshofer:2012ur, Giudice:2012qq}.  
However, charm is also challenging for theory because of the intermediate mass of the charm quark compared to the QCD scale.
Charm is often considered insensitive to short-distance physics, because of unknown, potentially large effects from QCD in its nonperturbative regime. 
As it is challenging to calculate nonleptonic charm decays from first principles, the methodology is often based
on flavor symmetry methods, namely the approximate SU(3)-flavor symmetry of QCD. 
At leading order, the usefulness of SU(3)-flavor symmetry is limited as it comes with corrections of order $\mathcal{O}(30\%)$.
However, new theoretical methods~\cite{Gavrilova:2022hbx} allow to take into account higher-order corrections in a systematic way in form of higher-order sum rules. These remain valid at much higher precision, depending on the order of the expansion. Thus, combined with the increased precision of future measurements show solid potential for non-SM searches. 
Additional promising directions for the future are further application of the LCSR formalism~\cite{Khodjamirian:2017zdu}, the systematic treatment of final-state interactions in QCD-based models~\cite{Bediaga:2022sxw, Schacht:2021jaz} as well as long-term advancements by lattice QCD~\cite{Hansen:2012tf, USQCD:2019hyg, Boyle:2022uba}.
On top of BSM searches, charm decays can also be used in order to learn more about QCD in regions of the dynamics where calculations are not possible. 

Taking a step back and considering the development of the field of \CP violation in meson decays over time, historically, \CP violation in charm decays has been found late, almost 20 years after it has been found in $B$ decays. The reason behind that is that \CP violation in singly-Cabibbo suppressed charm decays is proportional to  the small non-unitary contribution of the relevant 2$\times$2 submatrix of the CKM matrix.
Thus becomes transparent if one compares the dominating dynamics of charm and $B$ decays that lead to \CP violation.
\CP violation in $B$ decays stems mainly from interference of amplitudes with tree operators with those from penguin operators. Here, the size of penguin amplitudes is determined mainly by the ratio of the top-quark and $W$-boson masses, $m_t/m_W>1$. 
In charm decays the situation is different, because the size of amplitudes from 
penguin operators is determined by the ratio of the bottom-quark and $W$-boson masses, $m_b/m_W\ll 1$. In addition, in charm decays the GIM cancellation is achieved very precisely. It follows that penguin operators are not relevant for charm decays and this suppresses \CP violation.

However, the above argument does not hold for penguin contractions of tree operators: these may be sizable due to nonperturbative effects from QCD, including rescattering.

Direct charm \CP violation has first been discovered in the combination of \CP asymmetries~\cite{LHCb:2019hro, Amhis:2019ckw}  
\begin{align}
\Delta a_{CP}^{dir} &\equiv a_{CP}^{dir}(D^0\rightarrow K^+K^-) - a_{CP}^{dir}(D^0\rightarrow \pi^+\pi^-) = (-0.161 \pm 0.028)\%\,.
\end{align}
and was followed, very recently, by the first evidence of direct \CP violation in a single decay~\cite{LHCb:2022vcc}
\begin{align}
a_{CP}^{dir}(D^0\rightarrow K^+K^-) &= (7.7\pm 5.7)\times 10^{-4}\,,\\
a_{CP}^{dir}(D^0\rightarrow \pi^+\pi^-) &= (23.2\pm 6.1) \times 10^{-4}\,.
\end{align}
and associated theoretical interpretations~\cite{Schacht:2022kuj, Wang:2022nbm}. 

In the SM, we have schematically 
\begin{align}
\Delta a_{CP}^{dir} &\sim 10^{-3} \times r_{\mathrm{QCD}}\,,
\end{align}
with $r_{\mathrm{QCD}}$ being the ratio of pure low-energy QCD amplitudes. In the language of flavor-symmetries of QCD, this is the ratio of $\Delta U=0$ over $\Delta U=1$ $U$-spin matrix elements. Depending on the theory scenario for $r_{\mathrm{QCD}}$, significantly different interpretations of the data emerge, as demonstrated by the variety of methodologies discussed in  Refs.~\cite{Grossman:2019xcj, Brod:2011re, Soni:2019xko, Schacht:2021jaz, Khodjamirian:2017zdu, Chala:2019fdb, Li:2019hho, Bediaga:2022sxw}. 

In the future, it will be crucial to be able to assess which of the methods give the correct value of $r_{\mathrm{QCD}}$.
One important direction for future research is therefore to confront the predictions of these methods with additional charm decays, 
most notably with the $D\rightarrow \pi\pi$ system. The latter is sensitive to the ratio of $\Delta I=1/2$ over $\Delta I=3/2$ isospin matrix elements, which can be extracted from experimental data to be $\mathcal{O}(1)$~\cite{Franco:2012ck}.
In Ref.~\cite{Grossman:2019xcj} it is claimed that it would be natural that $r_{\mathrm{QCD}}$ would follow a similar pattern,~\emph{i.e.},~$r_{\mathrm{QCD}}\approx 1$, which is the so-called $\Delta U=0$ rule.  In the light cone sum rule (LCSR) formalism on the other hand it is found $r_{\mathrm{QCD}}\approx 0.1$~\cite{Khodjamirian:2017zdu}. It would therefore be extremely interesting to work out the LCSR predictions for the charm $\Delta I=1/2$ rule and to see if LCSRs can reproduce the $\mathcal{O}(1)$ ratio here while giving $r_{\mathrm{QCD}}\approx 0.1$ at the same time. Another crucial direction for future research are higher-order sum rules for observables, especially for \CP asymmetries.

\subsection{Experimental overview}

Measurements of \CP violation in bottom and charm processes span a large variety of different experimental approaches that range from relatively straightforward determinations of charge-dependent yield asymmetries in charged hadron decays to sophisticated decay-time-dependent yield asymmetries between flavor-tagged meson decays identified over the Dalitz plot space. However, the reach in precision is typically dependent on a restricted set of key performance drivers.  Reconstruction of large and low-background decay samples is of paramount importance. Production cross-sections for $pp\to b\bar{b}$ and $pp\to c\bar{c}$ processes of 100--1000$\mub$ offer LHC experiments large advantage in raw production rate with respect to the 1 nb cross sections at energy-asymmetric electron-positron colliders at the $\Upsilon(4S)$ resonance (so-called $B$-factories).  In addition, all species of flavored hadrons are produced in hadron collisions, including bottom-strange mesons and bottom baryons, which are kinematically prohibited at the $B$-factories.
However, online event-selections typically based on particle transverse-momentum and displacement from the interaction point are needed to control backgrounds 1000-fold larger than signals in hadron collisions. While these reduce the raw signal-yield difference, significant, 10--1000-fold, advantages remain for experiments at hadron colliders. Moreover, ``triggerless" online-selection architectures such as those deployed at the LHCb experiment in 2022 allow for relaxing the momentum and displacement requirements thus recovering a larger fraction of the raw yield. In general, the yield advantage is mostly exploited in final states including only charged particles, where the precision of position-sensitive charged-particle detectors allow for discriminating signals from the enormous backgrounds. Efficient reconstruction of final states involving neutral particles or neutrinos is challenging. In addition, difficulties in modeling the complicated biases induced by stringent online event-selections and flavor asymmetries at production induced by asymmetric polar coverage make absolute measurements challenging,
In spite of comparatively lower yields and fewer accessible initial states, the Belle II experiment has unique advantages over hadron-collider experiments. Backgrounds at production are typically 100-fold smaller due to the absence of collision pile-up and high discriminating power of simple track multiplicity and total event-energy criteria that fully suppress high-rate backgrounds from electroweak processes. The rate of remaining background from ``continuum'' production of light-quark pairs is only 3-4 times higher than signal rates natively, and is further suppressed using the particle distributions in the detector volume, which are significantly distinctive owing to at-threshold production.  $B$-factory collisions produce $B$ meson pairs with no additional particles.  Reconstruction of properties of neutral particle (photon, and neutral pions and kaons) is nearly as efficient and precise as that of charged particles. Because the initial state is known and the detector is nearly hermetic, $B$-factory experiments reconstruct fully-inclusive final states and broadly search for particles with little or no direct signature in the detector. Reconstruction efficiencies in $B$-factory experiments are largely uniform in decay time and kinematic properties of the final states, which offers an advantage in measurements that depend critically on the description of multiparticle signal kinematics, as in Dalitz plot analyses. Higher-level, ``flavor-tagging" algorithms that reconstruct the particles accompanying the signal identify the flavor of neutral $B$ mesons for decay-time-dependent \CP-violating yield asymmetries. Coherent $B^0\overline{B}^0$ pair production at the $B$-factories allows identifying the flavor in 30\% of the signal candidates, whereas only a 5\% of the signal candidates is typically flavor-tagged in hadron collisions. \par Currently, \CP violation has been established (at significances greater than five standard deviations) in several bottom and charm observables~\cite{Workman:2022ynf}.

\section{Specific $B$ decay modes}

\subsection{$B \to DK$}

The quark-mixing parameter 
$\gamma/\phi_3 \equiv \arg (-V_{ud}V^{*}_{ub}/V_{cd}V^{*}_{cb})$, 
where $V_{qq'}$ are CKM matrix elements, is measurable via the interference of the tree-level quark transitions 
$\bar{b} \to \bar{c}u\bar{s}$ and  $\bar{b} \to \bar{u}c\bar{s}$ 
accessible through \CP violation in $B^+ \to D K^+$ decays, 
where $D$ is either a $D^0$ or 
a $\overline{D}^0$~\cite{Carter:1980hr,Carter:1980tk, Gronau:1991dp, Gronau:1990ra, Atwood:1996ci, Atwood:2000ck, Bondar:2002, Giri:2003ty, Belle:2004bbr, Ceccucci:2020cim}. The tree-level nature of these transitions results in negligible theoretical uncertainties when interpreting the measured observables. Therefore, assuming the absence of non-SM contributions at tree level, the measurement of $\gamma$ provides a test of the SM when compared to indirect determinations.
Yet, with future data, more sophisticated analysis methods will be needed in order to take into account subleading effects such as $D^0-\overline{D}^0$ mixing. Other opportunities for improvement may come from optimized usage of the available data.  For example, optimization of binning of the $D \to K^0_S h^+h^-$ Dalitz space is currently based on a specific model.  It is unclear, however, if better binnings could be available and if they can be adjusted to be optimal for the available charm data.  Alternatively, the use of unbinned methods may be exploited~\cite{Poluektov:2017zxp}. The advantage is avoiding loss of statistical sensitivity associated with binning even though impact of backgrounds and experimental effects on systematic uncertainties might be nontrivial. A synergy between additional theoretical work and experimental application on data is likely to offer future validation and optimizations of these methods. One other direction in which theory can bring about improvements is the exploration of optimal approaches to combine the various measurements. One outstanding question concerns whether we can use isospin symmetry to combine some measurements and reduce the number of parameters or not. While using isospin introduces a theoretical uncertainty, using isospin is likely to still reduce the total uncertainty as long as the experimental uncertainty is at the few-percent level. 
The current global precision of $3^\circ - 4^\circ $ is dominated by measurements based on $B^+ \to D(\to K^0_S \pi^+\pi^-) K^+$ decays reported by the LHCb experiment. These are expected to drive the precision over the next few years until Belle II will also become competitive with its full  50 ab$^{-1}$ sample. Asymptotically, both experiments will reach $1^\circ$ precision or better. Similar precision from experiments affected by largely different systematic uncertainties offers complementarity and redundancy, which are crucial when establishing the value of a fundamental
parameter that has deep implications on our understanding of SM \CP violation.

\subsection{$B \to J/\psi K^0$ and related modes}

Decay-time-dependent decay-rate asymmetries between particles and antiparticles offer multiple probes of contributions from massive non-SM particles to the mixing or decay amplitudes. The goal is to test for significant discrepancies between observed asymmetries and asymmetries predicted by the CKM hierarchy, or between asymmetries observed in different channels dominated by the same SM phases.
The decay modes $B^0\rightarrow J/\psi K^0$ and $B^0_s\rightarrow J/\psi \phi$, governed by the parton-level process $\bar{b}\rightarrow \bar{c}c\bar{s}$, are the most effective probes of \CP violation arising from the interference between mixing and decay in $B^0$ and $B^0_s$ mesons~\cite{Bigi:1981qs}. These decays are dominated by a single tree amplitude that cancels to good precision in the determination of the decay-time-dependent decay-rate asymmetry between particle and antiparticles, which is therefore directly proportional to $\sin(2 \beta_{(s)})$. 
However, this approach relies on the assumption that subleading effects,  most notably the contributions from penguin diagrams, are sufficiently suppressed. While this is justified in practice for the current $\mathcal{O}(0.01)$ precision, so-called penguin-pollution effects have to be taken into account for precision beyond the percent level. Similar considerations hold for the effects of kaon mixing~\cite{Grossman:2002bu}. Determination of these subleading effects is the key for more precision.  Methodologies for treating the subleading effects are laid out in 
Refs.~\cite{Fleischer:1999nz, Boos:2004xp, Ciuchini:2005mg, Beneke:2005pu, Li:2006vq, Gronau:2008cc, Faller:2008gt, Faller:2008zc, Ciuchini:2011kd, Barel:2020jvf, Jung:2012mp, DeBruyn:2014oga, Ligeti:2015yma, Frings:2015eva}.
These works can be categorized into ones using flavor symmetry relations~\cite{Fleischer:1999nz, Ciuchini:2005mg, Faller:2008gt, Faller:2008zc, Ciuchini:2011kd, Barel:2020jvf, Jung:2012mp, DeBruyn:2014oga, Ligeti:2015yma}, rescattering models~\cite{Gronau:2008cc}, 
or calculations based on QCD factorization techniques
\cite{Boos:2004xp, Beneke:2005pu, Li:2006vq, Frings:2015eva}. It is likely that auxiliary experimental inputs from other decays that are specifically sensitive to similar penguin amplitudes will mitigate the challenges associated with calculating theoretically penguin contributions.

Using tree-dominated $(c\bar{c})K^0$ final states, first-generation $B$-factory experiments
and LHCb achieved determinations of $\beta/\phi_1\equiv \arg(-V_{cd}V^{*}_{cb}/V_{td}V^{*}_{tb})$ at 2.4\% precision, with systematic uncertainties typically associated with limited knowledge of 
the vertex reconstruction model and flavor tagging. This offers a reliable and
precise SM reference whose precision is expected to further improve to below 1\% in the
next decade thanks to both LHCb and Belle II.  An important priority is therefore to approach that precision in the corresponding modes governed by loop amplitudes to probe any discrepancies. 
The highest priority channel, $B^0\to \eta' K^0$, has a sizable decay rate dominated by the $b \to s$ loop amplitude. Similarly promising are the 
channels $B^0 \to \phi K^0$, despite challenges
associated with model-related systematic uncertainties from the Dalitz-plot analysis, and the processes $B^0 \to K^0\pi^0\gamma$, $K^0 \pi^+\pi^-\gamma$, and $\rho^0 \gamma$.
Since all these channels involve final-state $\pi^0$, $K^0$, or $\gamma$ Belle II offers the most promising opportunities with expected precisions on $\sin(2\beta_{\rm eff})$ of up to 10\% for $B^0\to \eta' K^0$ or even 5\% for $B^0 \to K^0\pi^0\gamma$. 

After the first pioneering Tevatron measurements, LHCb and, to a lesser extent CMS and ATLAS, dominate the current $\sin(2\beta_s)$ precision using tree-dominated $B^0_s \to J/\psi \phi$ decays. Improvements of current constraints will continue to come from LHC experiments, and further improved with LHCb's loop-dominated $B^0_s \to \phi\phi$ and $B^0_s \to K^{*0}\overline{K}^{*0}$ results.

\subsection{$B \to \pi \pi$ and related modes}

Studies of $B$ decays with no charm quark in the final state, so-called charmless decays,  give access to
$\alpha/\phi_2 \equiv \arg[-V^{*}_{tb}V_{td}/V^{*}_{ub}V_{ud}]$, the least known angle of the CKM unitarity triangle, and probe non-SM physics in
processes mediated by loop decay-amplitudes. However, an unambiguous interpretation of the
results is spoiled by hadronic uncertainties, which are hardly tractable in perturbative calculations. Appropriate combinations of measurements from decays related by flavor symmetries reduce the impact of such unknowns.
The underlying quark-level transitions of charmless $B$ decays are the tree-level $b\rightarrow u$ and loop-level $b\rightarrow d,s$ transitions. The top-quark loop plays an important role in the latter because $m_t\gg m_W$. 
An important characteristic of $b\rightarrow u\bar{u} d$ decays is that penguin and tree processes share the same order in the Wolfenstein-$\lambda$ expansion, but have a different weak phase. The decays $B\rightarrow \pi\pi$, $B\rightarrow \rho\rho$, and $B\rightarrow \pi\rho$ play an important role for the extraction of the CKM angle~$\alpha$.
To suppress the uncertainty introduced by challenging hadronic amplitudes, the isospin analysis is crucial~\cite{Gronau:1990ka}. Isospin symmetry is used to define a sum rule between the amplitudes of $B^0\rightarrow \pi^+\pi^+$, $B^+\rightarrow \pi^+\pi^0$ and $ B^0\rightarrow \pi^0\pi^0$ as well as for the corresponding triangle construction. Charmless $B$ decays can provide constraints on the CKM angle~$\gamma$~\cite{Paz:2002ev, Gronau:1994bn, Gronau:1994rj, Fleischer:1997um, Neubert:1998pt, Gronau:1998fn, Gronau:1999qd}.
Here, an important ingredient is the relation between electroweak penguin and tree diagrams found in Ref.~\cite{Neubert:1998pt}.
Note that although no charm quark appears in the final state, the charm quark still plays an important role in terms of \lq\lq{}charming penguin\rq\rq{} diagrams~\cite{Ciuchini:1997hb}. 

In addition to flavor-symmetry based methods~\cite{Gronau:1995hm, Gronau:1994rj, Zeppenfeld:1980ex}, key phenomenological tools 
are the heavy quark expansion and QCD factorization (QCDF)~\cite{Beneke:1999br, Beneke:2000ry, Beneke:2001ev, Beneke:2003zv, Grossman:2005jb, Lu:2022fgz} as well as soft collinear effective field theory (SCET)~\cite{Bauer:2001yt, Bauer:2004tj, Bauer:2005kd, Arnesen:2006vb,  Williamson:2006hb}. Other methodologies in the literature are the factorization-assisted topological approach~\cite{Zhou:2016jkv, Wang:2017hxe} and perturbative QCD~\cite{Keum:2000wi, Keum:2000ph, Lu:2000em, Ali:2007ff}. All of these approaches try to explore different paths beyond the picture of \lq\lq{}naive factorization\rq\rq{}~\cite{Bauer:1986bm}.
Naturally, a key question for the interpretation of the data is how big non-factorizable contributions are, see e.g., Ref.~\cite{Feldmann:2004mg}.
Attempts at matching the different parametrizations onto each other can be found in Refs.~\cite{He:2018joe, He:2018php, Huber:2021cgk}. 

Studies that explore the non-SM sensitivity of charmless $B$ decays can be found for example in Refs.~\cite{Ciuchini:2012gd, Bobeth:2014rra,  Fleischer:2018bld}, exemplifying that nonleptonic decays are a tool to learn more about both non-SM physics  and low-energy QCD.
Regarding the latter, an interesting question is how the analogy of the $\Delta I=1/2$ rule in kaons is reflected in $B$ decays, which is studied in Refs.~\cite{Grinstein:2014aza, Nayak:2020oba}.

Thanks to its excellent reconstruction efficiencies and resolutions for neutral particles, Belle~II has the unique capability of studying jointly, and within the same experimental environment, all relevant final states of isospin-related charmless decays. Belle~II is therefore in a unique position for determining $\alpha/\phi_2$ and for testing the SM through isospin sum-rules at unprecedented precision.
Using $B^0 \to \rho \rho$ decays only, Belle II expects to improve the current global precision of 4 degrees on $\alpha$ down to 2 degrees. This will further halve, or better, if $B^0 \to \pi\pi$ and $B^0 \to \rho\pi$ decays will be used too; if the understanding of the contributions from poorly-known decays involving $a_1$, $f_0$, and non-resonant $\pi\pi$ final states will improve; and if  isospin-breaking effects will be constrained. LHCb's contributions to measurements involving final states with charged particles only will be precious. Belle~II will also strongly improve our knowledge of the branching fraction and direct \CP-violating asymmetry of the $B^0 \to \pi^0\pi^0$ decay expecting up to a ten-fold improvement. 

\subsection{Baryonic decay modes}

Charmless multibody decays of $b$-baryons are expected to exhibit direct \CP violation, arising from the interference of charged-current $b\to u$ transitions with neutral-current $b\to d,s$ transitions.  These decays feature a rich underlying resonance structure in the two- and three-body invariant mass spectra of the final state constituents, which can potentially lead to large \CP-violating effects. 
LHCb reported the first evidence for baryonic \CP violation in the decay $\Lambda_b^0\to p \pi^- \pi^+\pi^-$~\cite{LHCb:2016yco} using triple products of final-state particle momenta to construct \CP-violating observables as functions of phase-space. One kinematic region exhibited a significant \CP-violating signal, more than $3\sigma$ away from zero. Measurements of kinematic-integrated \CP-violating asymmetries in  $X^0_b\to p h h^\prime h^{\prime\prime}$ decays, where $X^0_b=\Lambda_b^0,\Xi_b^0$, and $h,h^\prime, h^{\prime\prime}$ are pions or kaons, show no significant \CP violation~\cite{LHCb:2019jyj}.
These results, along with recent lattice results on form factors of baryons~\cite{Detmold:2015aaa} are promising; however, the treatment of non-leptonic baryon decays remains a challenge. LCSR results for baryon form factors are given in Refs.~\cite{Wang:2008sm, Khodjamirian:2011jp, Wang:2015ndk}.
pQCD predictions for $\Lambda_b$ decays can be found in Ref.~\cite{Lu:2009cm}.  

It is an open question if the methodology used for meson decays works equally well for baryon decays.
Yet, a lot of opportunities, \emph{e.g.}, for SM tests with sum rules based on flavor symmetries lie in front of us to probe in the next decade, see \emph{e.g.},~Refs.~\cite{Dery:2020lbc, Roy:2020nyx}. Testing if the anomalies observed in meson decays are seen also in baryon decays governed by the same underlying quark-level processes is a key consistency check. On top of that, baryon decays allow to probe many additional observables due to spin correlations.

\section{Specific $D$ decay modes}

\subsection{Direct \CP violation in singly-Cabibbo-suppressed decays}

After a multiyear effort by several experiments~\cite{E791:1997txw, FOCUS:2000ejh, CLEO:2001lgl, BaBar:2007tfw, Belle:2008ddg, CDF:2011ejf, CDF:2012ous, LHCb:2011osy, LHCb:2013dkm, LHCb:2014kcb, LHCb:2016csn, LHCb:2016nxk}, \CP violation has been observed in charm for the first time, by the LHCb experiment~\cite{LHCb:2019hro}. This is just 
the beginning of the exploration of \CP violation phenomena in charm decays. In the future it will be crucial to measure the direct \CP-violating asymmetries of all singly-Cabibbo-suppressed two-body charm decays. This is necessary in order to benefit from the insights of flavor symmetries, which relate reliably the various \CP asymmetries in terms of sum rules.  The most promising channels in the near future are $D\rightarrow K^0_SK^0_S$~\cite{Nierste:2015zra, Hiller:2012xm, Atwood:2012ac} and $D\rightarrow KK^*$~\cite{Nierste:2017cua}. 

In order to test if our picture of the breaking of flavor symmetries is correct, we need to test the SU(3)$_F$-limit sum rules~\cite{Grossman:2006jg, Pirtskhalava:2011va, Grossman:2012ry, Grossman:2013lya, Muller:2015rna}
\begin{align}
a_{CP}^{\mathrm{dir}}(D^0\rightarrow \pi^+\pi^-)+a_{CP}^{\mathrm{dir}}(D^0\rightarrow K^+K^-) &= 0\,, \\
a_{CP}^{\mathrm{dir}}(D^+\rightarrow K^0_S K^+) + a_{CP}^{\mathrm{dir}}(D_s^+\rightarrow K^0_S\pi^+) &= 0\,,
\end{align}
and check if their breaking is at the expected 30\% level. More precise sum rules that account for additional sources of SU(3)$_F$-breaking can be obtained when including an additional \CP asymmetry. These rules test the relation among ~\cite{Muller:2015rna}
\begin{align}
a_{CP}^{\mathrm{dir}}(D^0\rightarrow \pi^+\pi^-)\,, \quad
a_{CP}^{\mathrm{dir}}(D^0\rightarrow K^+K^-)\,, \quad
a_{CP}^{\mathrm{dir}}(D^0\rightarrow \pi^0\pi^0)\,,
\end{align}
and
\begin{align}
a_{CP}^{\mathrm{dir}}(D^+\rightarrow K_S K^+)\,, \quad a_{CP}^{\mathrm{dir}}(D_s^+\rightarrow K_S\pi^+)\,, \quad 
a_{CP}^{\mathrm{dir}}(D_s^+\rightarrow K^+\pi^0)\,,
\end{align}
respectively.
Many more sum rules exist for more complicated systems, like baryon decays and decays to vector mesons~\cite{Altarelli:1975ye, Kingsley:1975fe, Korner:1978tc, Matsuda:1978cj, Savage:1989qr, Pakvasa:1990if, Savage:1991wt, Verma:1995dk, Sharma:1996sc, Kang:2010td, Jia:2019zxi} such as~\cite{Grossman:2018ptn}
\begin{align}
a_{CP}^{\mathrm{dir}}(\Lambda_c^+\rightarrow p K^-K^+) + a_{CP}^{\mathrm{dir}}(\Xi_c^+\rightarrow \Sigma^+\pi^-\pi^+) &= 0\,,\\
a_{CP}^{\mathrm{dir}}(\Lambda_c^+\rightarrow \Sigma^+ \pi^- K^+ ) + a_{CP}^{\mathrm{dir}}(\Xi_c^+\rightarrow p K^- \pi^+) &= 0\,, \\
a_{CP}^{\mathrm{dir}}(\Lambda_c^+\rightarrow p \pi^- \pi^+ ) + a_{CP}^{\mathrm{dir}}(\Xi_c^+\rightarrow \Sigma^+ K^-K^+)&= 0\,.
\end{align}
In addition to flavor sum rules, it is important to scrutinize 
the $\Delta U=0$ rule further, for example by testing if the same pattern can also be seen in three-body decays~\cite{Dery:2021mll}. Three-body decays have the advantage of enabling the extraction of the relative phase between $\Delta U=0$ and $\Delta U=1$ matrix elements already with time-integrated measurements~\cite{Dery:2021mll},
due to interference effects in the Dalitz plot whereas two-body decays require decay-time-dependent or quantum-correlated measurements. 
Measurements of the $D\rightarrow \pi\pi$ system can be used to study the $\Delta I=1/2$ and $\Delta I=3/2$ matrix elements~\cite{Grossman:2012eb}. This offers a powerful SM null test~\cite{Buccella:1992sg, Grossman:2012eb} prescribing 
\begin{align}
a_{CP}^{dir}(D^+\rightarrow \pi^+\pi^0)=0\,,
\end{align}
which holds in the isospin limit.
In the future, it will be especially important to identify observables optimized for maximizing the sensitivity to \CP violation in multibody charm decays. An important role could be played by triple-product asymmetries~\cite{Durieux:2015zwa}. Open questions in this area include the following:  What is the most sensitive binning of the phase space for multibody charm decays? And how exactly can we formally account for the phase space effects when comparing Dalitz plots that are related by flavor symmetries?
Without an answer to the last question it will hardly be possible to improve the theoretical precision for multi-body decays beyond the generic estimate of $30\%$ SU(3)$_F$ breaking.
Another important direction for non-SM physics searches is the continued search for \CP violation in doubly-Cabibbo suppressed decays, where special care has to be taken to account for contributions from kaon \CP violation~\cite{Yu:2017oky}, and Cabibbo-favored decays.
If future measurements turn out to be in tension with the SM interpretation as formulated in the $\Delta U=0$ rule, that would strengthen the case for the existence of non-SM physics with flavor non-diagonal couplings to the up sector, in order to explain the measured $\Delta a_{CP}^{dir}$. Candidate benchmark models include a 2HDM where scalar $(\bar u c)$ and $(\bar u u)$ couplings exist, and models with vector-like up-quarks inducing $(\bar u c)$ $Z$ couplings. The scale of non-SM physics required by these models is model dependent, and ranges from hundreds of GeV to a few tens of TeV. 

Given the prominent role that flavor symmetries have in this program, the synergetic interplay between Belle II and LHCb will be essential to extract the maximum information from the data. While the capability of LHCb to improve the current subpercent precision in final states with only charged particles will depend on the understanding of systematic uncertainties, Belle II's challenge will be to gather sufficient data to approach those precisions in final states with $\pi^0$  mesons.

\subsection{Mixing and decay-time-dependent charm \CP violation}

Recently, important new measurements of charm mixing have been achieved. A nonzero value for the mixing parameter $x$, which measures the $D$ meson mass splitting has now been observed at more than $7 \sigma$~\cite{LHCb:2021ykz}. The uncertainty of $y$, which measures the difference of the neutral $D$ decay widths, has been reduced by a factor two~\cite{LHCb:2021dcr}. The calculation of charm-mixing parameters in the SM is challenging, see Ref.~\cite{Lenz:2020efu} for recent progress,
where the heavy quark expansion (HQE) result for $y$ has been found in agreement with the experimental value. In order to bring the large theory error under control, future research into higher-order HQE contributions is important~\cite{Lenz:2020awd}. Another theoretical approach to the calculation of charm mixing is the insertion of a complete set of hadronic states into the mixing amplitude of $D^0$ and $\overline{D}^0$~\cite{Wolfenstein:1985ft, Donoghue:1985hh, Falk:2001hx, Cheng:2010rv}.
Recently, there has also been progress from lattice QCD~\cite{Bazavov:2017weg, Carrasco:2015pra, USQCD:2019hyg, Boyle:2022uba}.  Altogether, to date we have qualitative agreement of charm mixing parameters with the SM.  Mixing is also related to decay-time-dependent measurements of \CP violation in charm. The latter are crucial for singly-Cabibbo-suppressed $D\rightarrow PP'$ decays, as --  quantum-correlated charm-measurements aside -- they offer the only access to the strong phase between the relevant  CKM-suppressed and CKM-leading amplitude~\cite{Grossman:2019xcj}.
For example, for the extraction of the $\Delta U=0$ rule, \emph{i.e.},~$r_{\mathrm{QCD}}\sim 1$, from the data,  it is necessary to assume that the relative strong phase between the $\Delta U=0$ and $\Delta U=1$ matrix elements has a generic value of $\mathcal{O}(1)$. However, this phase could be accidentally suppressed, which would change the interpretation of the  measurement of $\Delta a_{CP}^{\mathrm{dir}}$. Decay-time-dependent measurements are therefore a unique way to shed more light onto this important question, and to completely solve the $U$-spin system of charm decays for the underlying theory parameters. This makes the search for time-dependent \CP violation in charm decays key both for understanding better the SU(3)-flavor anatomy of QCD and for the quest for non-SM physics. 

Here the large charm hadroproduction rate offers unsurpassed sensitivity of the LHCb experiment. Belle~II measurements might contribute whenever complementary information from decays into final states including a $\pi^0$ will be important. 

\section{Summary}
Since \CP is not a fundamental symmetry of nature, non-Standard-Model particles generically appear with order-one \CP-violating phases, leading  potentially to observable and large deviations from some Standard Model predictions. Thus, the exploration of \CP violation offers multiple avenues for searching for non-Standard-Model particles, at masses that are orders of magnitude higher than those directly accessible in high-energy hadron collisions. 
We briefly review the current status of \CP-violation studies in bottom and charm physics focusing on those quantities that are most sensitive to non-Standard-Model contributions. We argue that the current simultaneous operation of two dedicated experiments exploiting different collision environments and complementary experimental capabilities offers an unique and compelling opportunity to pursue this program. We believe that the study of \CP violation in bottom and charm decays by the Belle~II and LHC experiments as well as the corresponding theoretical advancements primarily deserve to be among the top HEP priorities for the next decade and beyond.

\section*{Acknowledgments}
The work of Y.G. is supported in part by the NSF grant PHY1316222. S.S. is supported by a Stephen Hawking Fellowship from UKRI under reference EP/T01623X/1 and the Lancaster-Manchester-Sheffield Consortium for Fundamental Physics, under STFC research grant ST/T001038/1.

\setboolean{inbibliography}{true}
\addcontentsline{toc}{section}{References}
\bibliographystyle{LHCb}
\bibliography{references}
\setboolean{inbibliography}{false}

\end{document}